\documentstyle[12pt,psfig,epsfig,aaspp4]{article}
\def\degr{\hbox{$^\circ$}}
\def\arcmin{\hbox{$^\prime$}}

\def\fdg{\hbox{$.\!\!^\circ$}}
\def\farcm{\hbox{$.\mkern-4mu^\prime$}}
  
\def\lsim{\mathrel{\hbox{\rlap{\lower.55ex \hbox {$\sim$}}\kern-.0em
\raise.4ex \hbox{$<$}}}} 
\def\gsim{\mathrel{\hbox{\rlap{\lower.55ex \hbox {$\sim$}}\kern-.0em
\raise.4ex \hbox{$>$}}}}

\begin{document}

\title{The 1.4 GHz light curve of GRB~970508}
\author{T.J. Galama\altaffilmark{1}, R.A.M.J. Wijers\altaffilmark{2},
M. Bremer\altaffilmark{3}, P.J. Groot\altaffilmark{1}, 
R.G. Strom\altaffilmark{1,4}, A.G. de Bruyn\altaffilmark{4,5},
C. Kouveliotou\altaffilmark{6,7}, C.R. Robinson\altaffilmark{6,7},
J. van Paradijs\altaffilmark{1,8}} 
\altaffiltext{1}{Astronomical Institute `Anton Pannekoek', University of Amsterdam,
\& Center for High Energy Astrophysics,
Kruislaan 403, 1098 SJ Amsterdam, The Netherlands}
\altaffiltext{2}{Institute of Astronomy, Madingley Road, Cambridge, UK}
\altaffiltext{3}{Institut de Radio Astronomie Millim\'{e}trique,
300 rue de la Piscine, 
F--38406 Saint-Martin d'H\`{e}res, France}
\altaffiltext{4}{NFRA, Postbus 2, 7990 AA Dwingeloo, The Netherlands}
\altaffiltext{5}{Kapteyn Astronomical Institute, Postbus 800, 9700 AV,
Groningen, The Netherlands} 
\altaffiltext{6}{Universities Space Research Asociation}
\altaffiltext{7}{NASA/MSFC, Code ES-84, Huntsville AL 35812, USA}
\altaffiltext{8}{Physics Department, University of Alabama in
Huntsville, Huntsville AL 35899, USA}


\begin{abstract}
We report on Westerbork 1.4 GHz radio observations of the radio
counterpart to $\gamma$-ray burst GRB~970508, between 0.80 and 138
days after this event. The 1.4 GHz light curve shows a transition from
optically thick to thin emission between 39 and 54 days after the
event.  We derive the slope $p$ of the spectrum of injected electrons
(d$N$/d$\gamma_{\rm e}\propto\gamma_{\rm e}^{-p}$) in two independent
ways which yield values very close to 
$p=2.2$. This is in agreement with a relativistic dynamically
near-adiabatic blast wave model whose emission is dominated by
synchrotron radiation and in which a significant fraction of the
electrons cool fast.
\end{abstract} 



\section{Introduction}
The peak luminosities of $\gamma$-ray bursts (GRBs) are highly
super-Eddington and require relativistic outflows (Paczy\'{n}ski 1986;
Goodman 1986).  Paczy\'{n}ski and Rhoads (1993) pointed out that radio
emission is expected as a result of the interaction between such a
relativistic outflow and an external medium, as is, e.g., observed in
extragalactic jet sources (see also Katz 1994; M\'{e}sz\'{a}ros and
Rees 1997). They estimated that the strongest GRBs may be followed by
transient ($\sim 10$ mJy) radio emission at intervals ranging from
minutes (for a distance $d$ $\sim$ 10$^{5}$ pc) to several weeks ($d$
$\sim$ 10$^{9}$ pc).  However, searches for radio counterparts through
follow-up observations (Frail et al. 1994,1997a; Koranyi et al. 1995;
Galama et al. 1997a,b) were without success until recently.  With the
rapid accurate location capability of the Wide Field Cameras (WFCs;
Jager et al. 1995) onboard the Italian-Dutch X-ray observatory
BeppoSAX (Piro et al. 1995) it has recently become possible to detect
fading X-ray, optical and radio counterparts to GRBs (Costa et
al. 1997a; Piro et al. 1997a; Groot et al. 1997a,b; Van Paradijs et
al. 1997; Galama et al. 1997c,1998a; Sahu et al. 1997; Bond 1997;
Metzger et al. 1997; Frail et al. 1997b,c; Bremer et al. 1998; Halpern
et al. 1997).  These
observations have settled the discussion on the GRB distance scale
(`galactic halo' versus `cosmological', see e.g. Fishman and Meegan
1995, Lamb 1995, Paczy\'nski 1995):  GRBs occur at Gpc
distances.

GRB 970508 is the first GRB to be detected in the radio
(Frail et al. 1997b,c); the radio source position coincides with that
of the optical 
(Bond 1997) and X-ray (Piro et al. 1997b) afterglow sources. 
Assuming that the variations of the source at
4.86 and 8.46 GHz are due to interstellar scintillation (ISS), their
damping with time is consistent with a highly relativistically
expanding shell passing a diameter of $\approx 3\mu$as (Frail et
al. 1997b). VLBI observations show that the source is
unresolved ($<$ 0.3 mas, Taylor et al. 1997).

We here report on the results of 1.4 GHz radio observations of GRB 970508, made
with the Westerbork Synthesis Radio Telescope (WSRT) between 0.80 and
138 days after the burst occurred.

\section{Radio Observations}

On May 8.904 UT BeppoSAX recorded a moderately bright GRB
(Costa et al. 1997b) with the Gamma-Ray Burst  
Monitor (GRBM; Frontera et al. 1991), which was also recorded with the Wide 
Field Cameras on board BeppoSAX. 
Analysis of the WFC
observations gave a 3\arcmin\ (3$\sigma$ radius) error box centered on RA=
$06^{\rm h}53^{\rm m}28^{\rm s}$, Dec = +79\degr17\farcm4 (J2000; Heise et al.
1997b). The burst was also recorded (Kouveliotou et al. 1997) with the
Burst and  Transient Source Experiment (BATSE; Fishman et al. 1989) on
board the Compton Gamma-Ray Observatory. 

The error box of GRB~970508 was
first observed at 1.4 GHz with the WSRT at the preliminary position given by
Heise et al. (1997) on 
May 9.70 UT, starting 0.80 days after the event, for 9.0
hours. 
We used the standard WSRT 1.4 GHz receiving system and
continuum correlator, 
providing us with 5 bands of width 10 MHz and 3 bands of width 5 MHz. The
noise level in a continuum map after 12 hours of
integration is typically 0.05 mJy beam$^{-1}$ for 14 telescopes and
the full 65 MHz bandwidth. For the declination of GRB 970508, at
1.4 GHz, the synthesized beamwidth is $13^{''}\times13^{''}$ (full
width at half maximum; FWHM), and the
field of view is about 
0\fdg6. Table \ref{Obs} provides a log of the observations.

The data were analyzed using the NEWSTAR software 
package\footnote{http://www.nfra.nl:80/newstar/}.  
The interferometer complex visibilities were first examined for
possible electromagnetic interference 
and other obvious defects. Bands with strong interference
were either deleted or carefully edited. Interference was usually
limited to about 5 \% of the data. The interferometer gain and
phase, for each observation and each band, were calibrated using the
standard WSRT calibrators 3C 48, 3C 147 and 3C 286 (15.96, 9.50 and
14.77 Jy, respectively at 1.4 GHz; Baars et al.\ 1977). We constructed
a model of the GRB 970508 field (used
for deconvolution) that contains 95 point sources and 60 clean components,
(together about 100 sources) and is complete down to a level
of $\sim$ 0.45 mJy. No self
calibration was performed. 

\subsection{The light curve \label{sec:radio}}
\begin{figure}[ht]
\centerline{\psfig{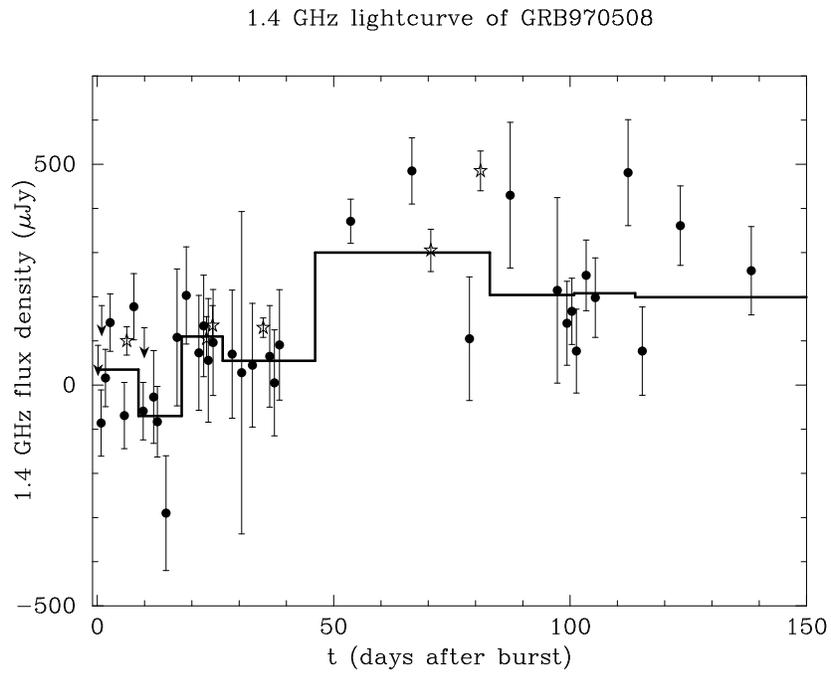}}
\caption[]{The 1.4 GHz light curve of GRB 970508. The data are from
Tab. \ref{Obs} ($\bullet$) and Frail et al. (1997b;
$\star$ and the 2$\sigma$ upper limits). The histogram represents
long-term average WSRT 1.4 GHz flux densities from Tab. \ref{tab:hist}.
\label{fig:light}}  
\end{figure}

Due to noise the
location of a source in the radio map can shift by roughly 0.5
FWHM/SNR, where SNR is the
signal-to-noise 
ratio of the detection. For very low SNR
detections the shift can be fairly large. If we would take a peak
in the map near the location of GRB 970508 we would bias ourselves
systematically to higher flux densities (as we cannot discern whether
the peak is due to noise on top of the detection or the detection
itself). Therefore, we constructed the light curve using the
flux density at the 
pixel on the exact position of the radio counterpart (Frail 1997c). The
errors (1$\sigma$) are the 
r.m.s. map noises (determined by a quadratic fit to the deconvolved
image). The maps
are super sampled  with $\sim$ 6 pixels beam$^{-1}$.
Negative flux density values in the 1.4 GHz light curve are due to
r.m.s. noise fluctuations on a non- or weakly-detectable source. 
The 1.4 GHz light curve is shown in Fig. \ref{fig:light}. 
We also divided the data from May 9 until September 23 
into 8 parts and added the observations in each
part to determine average flux densities. In Table
\ref{tab:hist} and in Fig. \ref{fig:light} 
we present these average flux densities. 

During the first 40 days  after the
onset of GRB 970508 the 1.4 GHz radio counterpart is not detected. 
A combined map of all observations (May 9.7-June 16.44 UT; excluding
June 8.43 UT for reasons of data quality) yields 33 $\pm$ 40
$\mu$Jy. The average 4.86 GHz flux
density during the first 50 days is 560 $\pm$ 10
$\mu$Jy (Frail et al. 1997b), implying that, on
average, the spectral  
index $\alpha_{\rm 1.4-4.86 GHz} >$ 1.5
(where we used a 2$\sigma$ upper limit of 80 $\mu$Jy at 1.4 GHz),
i.e. the radiation is self-absorbed ($F_{\nu} \propto \nu^2$; Katz and
Piran 1997).
At later times ($t >$ 50 days) 1.4 GHz emission is detected at, on
average,  228 $\pm$ 30 $\mu$Jy (Jul 1.49--Sep 23.22; 53.6-138.3 days
after the event). From Tab. \ref{tab:hist} we see that then the spectral
indices $\alpha_{\rm 1.4-4.86 GHz}$ and $\alpha_{\rm 1.4-8.46 GHz}$ are
consistent with the expected low 
frequency tail of synchroton radiation ($F_{\nu} \propto \nu^{1/3}$;
Rybicky and Lightman 1979). Hence a transition from optically thick to thin
emission occurred between 39 and 54 days after the event (see also
Frail et al. 1997b).

During the first month  
the 8.46 and 4.86 GHz
flux densities show rapid fluctuations, attributed to ISS, with an ISS
diffractive time scale $t_{\rm dif}$ 
of less than a day, and decorrelation bandwidth $\nu_{\rm dec}$ of
less than 2 GHz (Frail et al. 1997b). For 1.4 GHz this implies $t_{\rm 
dif}$ $\sim$ 3 hours and $\nu_{\rm	 
dec}$ $\sim$ 0.8 MHz (Goodman 1997). Hence ISS is not likely to affect the 
1.4 GHz light curve: it was constructed from observations with
long integration times ($>$ 3
hrs) and the full
bandwidth  (65 MHz). 
At 1.4 GHz the critical
angular size below which diffractive scintillation can be
observed  is a factor 4.5 smaller than at 4.86 GHz (Goodman
1997). Assuming that the 
apparent expansion velocity of the blast wave is constant and using
the fact that
Frail et al. (1997b) 
observed scintillation  at 4.86 GHz until day $\sim$ 30
we would expect diffractive ISS to modulate the 1.4 GHz flux density
for $t < 7$ days. We have searched for diffractive 
ISS in the individual 5 and 10 MHz 
bands during the first two weeks, dividing  the data into segments of 
6 hours (a trade off between the ISS decorrelation time 
and a sufficient amount of data). We find only one $> 3\sigma$ detection
(550 $\pm$ 
160 $\mu$Jy; 3.4 $\sigma$) on May
11.51 at 1.415 MHz (10 MHz bandwidth). The probability of a 3.4
$\sigma$ detection in $\sim$ 100 independent
measurements (with an asumed Gaussian distribution) is $\sim$ 3 \%. We
conclude that ISS was not observed by us at 1.4 GHz.

\section{Discussion \label{Disc}}

The observed optical spectral slope $\alpha$ and the optical power law
decay of the light curve $F_{\nu} \propto t^{\delta}$ is not
consistent with the expected relation for the simplest blast wave
model ($\delta = 3\alpha/2$; e.g. Wijers, Rees and M\'esz\'aros
1997). The observed power law decay value, $\delta = -1.141 \pm 0.014$
($t > 2$ days, Galama et al. 1998b; see also Pedersen et al. 1998,
Castro-Tirado et al. 1998, Sokolov et al. 1998) would imply $\alpha =
-0.761 \pm 0.009$, while in the optical passband $\alpha = -1.12 \pm
0.04$ is observed (Galama et al. 1998c, from here on Paper II; see
also Sokolov et al. 1998).  In the following we show that this may be
explained by rapid cooling of a significant fraction of the electrons.

A population of electrons with a power-law distribution of
Lorentz factors $\gamma_{\rm e}$ (d$N$/d$\gamma_{\rm
e}\propto\gamma_{\rm e}^{-p}$)  
above some minimum value $\gamma_{\rm m}$ emits a power law
synchrotron spectrum above the frequency 
$\nu_{\rm m}$ (corresponding to radiating electrons with $\gamma_{\rm
m}$; e.g. Rybicki \& Lightman 1979).  Independently, above some Lorentz
factor $\gamma_{\rm c}$ 
the electrons may cool rapidly, and an extra break in the spectrum is expected
at the corresponding frequency $\nu_{\rm c}$
(Sari, Piran, \& Narayan 1998).
Beyond a certain time $t_0$ ($t_0$ is small $\sim$ 500 sec; see Paper II) 
the evolution of the blast wave is adiabatic (Sari et
al. 1998 and see Paper II); then
$\nu_{\rm m} < \nu_{\rm c}$, and the spectrum 
varies as $F_{\nu} \propto \nu^{-(p-1)/2}$ from $\nu_{\rm m}$ up to 
$\nu_{\rm c}$; above $\nu_{\rm   
c}$ it follows $F_{\nu} \propto \nu^{-p/2}$ and 
below $\nu_{\rm m}$ it follows  
the low frequency tail, $F_{\nu} \propto \nu^{1/3}$ (Sari et al. 1998). 
The evolution in time of the
GRB afterglow is determined by the evolution of these break
frequencies: $\nu_{\rm c} \propto t^{-1/2}$ and $\nu_{\rm m}
\propto t^{-3/2}$ (both decrease with time). 

The decay part of the optical R$_{\rm c}$ (Coussins R) band light
curve (in the optical passband 
$\nu > \nu_{\rm c}$ for $t\gsim$ 1.2 days; Paper II) goes as $F_{\nu}
\propto t^{(2-3p)/4}$, while the spectrum is then
$F_{\nu} \propto \nu^{-p/2}$ (Sari et al. 1998). This allows us to
make two independent 
measurements of $p$: using $F_{\rm R_{\rm c}}\propto t^{-1.141 \pm
0.014}$ we find $p= 2.188 \pm 0.019$ and using $\alpha_{\rm opt} =
-1.12 \pm 0.04$ gives $p = 2.24 \pm 0.08$.  The 
excellent agreement between the values of $p$ supports that a
significant fraction of the electrons cool rapidly and that the
evolution of the GRB remnant is adiabatic. Additional evidence for 
rapid cooling of a significant fraction of the electrons is given in
Paper II.

Observations by Bremer et al. (1998) with the IRAM
Plateau de Bure
Interferometer (PdBI) at 86 GHz show a maximum around $\sim$ 12
days. We identify this	 
maximum with the break frequency $\nu_{\rm m}$ passing 86 GHz at
$t_{\rm m,86 GHz} \sim$ 12 days (Paper
II). We expect, the 8.46 and 1.4~GHz emission 
to peak at $t_{\rm m,8.46 GHz} \sim$ 55
days and $t_{\rm m,1.4 GHz} \sim$ 180 days, respectively ($\nu_{\rm
m} \propto t^{-3/2}$). Near day 55, a shallow maximum can be seen 
in the 8.46 GHz light curve 
(Frail et al. 1997b). Unfortunately our 1.4 GHz light
curve cannot be used to test the presence of the maximum at that
frequency, both due to low  signal to noise and because it ends 150
days after the burst, i.e. before the predicted maximum.

Before $\nu_{\rm m}$ passes 8.46 GHz at $t_{\rm 
m,8.46 GHz}$ we expect the 8.46 GHz spectrum
to follow the low frequency tail $F_{\nu} \propto \nu^{1/3}$, while
after $t_{\rm m,8.46 GHz}$ it is expected to be $F_{\nu} \propto  
\nu^{-(p-1)/2} = \nu^{-0.6}$ (Sari et  
al. 1998 and we have used $p$ = 2.2). Thus, we predict  
a gradual transition between $t_{\rm m,8.46 GHz}
\sim$ 55 days and $t_{\rm m,4.86 GHz} \sim$ 80 days (when also at 4.86
GHz $\nu_{\rm m}$ has passed) from $\alpha$ = 1/3 to $\alpha = 
-0.6$. We note that this expectation is different from blast wave
models that do not include the effect of rapid cooling of a
significant fraction of the	 
electrons ($F_{\nu} \propto \nu^{-1.1}$ similar to the optical slope;
see e.g. Wijers et al. 1997). Also the decays at 8.46 GHz (after $t_{\rm
m,8.46 GHz} \sim$ 55 days) and 4.86 GHz (after $t_{\rm
m,4.86 GHz} \sim$ 80 days) are
expected to be different from that in the optical and X-ray passbands,
$F_{\nu} \propto t^{3(1-p)/4} = t^{-0.9}$; where we have used $p$ =
2.2). These predictions can 
be tested with the continued monitoring at the VLA at 4.86 and 8.46
GHz by Frail et al. (1998). 
 
The radio afterglow light curves of GRB 970508 (Frail et
al. 1997b and
this Letter) show a much more gradual evolution than expected (see
e.g. the fit to the 8.46 and 4.86 GHz data by Waxman, Kulkarni and
Frail 1998). Also a constant self-absorption frequency was expected
(e.g. Waxman et al. 1997) while we here show that a transition from
optically thick to thin emission occurred around $\sim$ 45 days.  For
$t < t_o$ Sari et al. (1998) predict a decrease with time of the
self-absorption frequency, $\nu_{\rm a}$, while for $t >$ $t_0$ the
self-aborption frequency $\nu_{\rm a}$ remains constant.  The
transition from optically thick to thin 1.4 GHz radiation then
suggests that $t_0$ $\sim$ 45 days. Also the 8.46 GHz light curve
(Frail et al. 1997b) suggests that
$t_0$ cannot be much smaller than 10 days, i.e. 10 days $\lsim$ $t_0$
$\lsim$ 55 days (we have extrapolated backwards in time from the 8.46
GHz peak at $t_{\rm m}$ $\sim$ 55 days with the expected dependence
$F_{\nu} \propto t^{1/2}$ for times $t < t_{\rm m}$). This is not in
agreement with the finding that $t_0 \sim 500$ sec (Paper II).
However, the absence of a break in the smooth
power law decay of the optical light curve from 2 to 60 days after the
burst (Pedersen et al. 1998; Castro-Tirado et al. 1998; Sokolov et
al. 1998; Galama et al. 1998c) shows that there is no important
transition in that period. This does imply that some
additional ingredient is needed; for example,
Waxman et al. (1998) argue that the transition from 
ultrarelativistic to mildly relativistic expansion of the blast wave may
explain the decrease in the self-absorption frequency $\nu_{\rm a}$
with time and the slow time dependence of the early radio light
curves.

The excellent agreement in the derived value for $p$ ($p$ = 2.2) from
the decay of	 
the optical light curve and the optical spectral slope support an
adiabatic dynamical 
evolution of the GRB remnant and an extra break in the synchrotron spectrum
at the frequency $\nu_{\rm c}$ above which the radiation is from electrons
which cool rapidly compared to the remnant's expansion time.
We predict a transition in the
radio spectral index $\alpha_{\rm 4.86-8.46 GHz}$ from 1/3 to --0.6, between  
55 and 80 days; the light curves are predicted to decay as 
$F_{\nu} \propto t^{-0.9}$ after 55 days
at 8.46 GHz and 80 days at 4.86 GHz. 

\acknowledgments

We are grateful for the assistance of the WSRT telescope operators
G. Kuper and R. de Haan. We would like to thank Dr. Frail for
communications on VLA observations of GRB 970508.
The WSRT is operated by 
the Netherlands Foundation for Research in Astronomy (NFRA) with
financial aid from the Netherlands Organization for Scientific Research
(NWO). T. Galama is supported through a grant from NFRA under contract
781.76.011. R. Wijers is supported by a Royal Society URF grant.
C. Kouveliotou acknowledges support from NASA grant NAG 5-2560.

\begin{table}
\caption[c]{1.4 GHz WSRT observations. \label{Obs}}
\begin{minipage}{8.8cm}
\begin{flushleft}
\begin{tabular}{lccc|lccc}
\tableline
UT day & $\Delta t$ & Length & $F_{\rm 1.4 GHz}$ & UT day & $\Delta t$ & Length & $F_{\rm 1.4 GHz}$\\
(1997)\footnote{Observing times refer to
the middle of the 
observing period}  & (days) & (hrs)& (mJy) & (1997)$^{a}$  & (days) & (hrs)& (mJy)\\
\tableline
May  9.70 &0.80		&  9.0 & --0.09	$\pm$ 0.08	&Jun 14.37 &36.47  	& 3.9  & 0.07 $\pm$ 0.12  \\  
May 10.64 &1.73		& 12.0 & 0.02 $\pm$ 0.07	&Jun 15.37 &37.47 	& 4.1  & 0.01 $\pm$ 0.12  \\ 
May 11.63 &2.73 	& 12.0 & 0.14 $\pm$ 0.07	&Jun 16.44 &38.54	& 7.4  & 0.09 $\pm$ 0.13  \\ 
May 14.62 &5.72 	& 12.0 & --0.07 $\pm$ 0.08 	&Jul 1.49  &53.59 	& 12.0 & 0.37 $\pm$ 0.05 \\   
May 16.63 &7.73		& 12.0 & 0.18 $\pm$ 0.08	&Jul 14.45 &66.55 	& 7.7  & 0.49 $\pm$ 0.08 \\    
May 18.61 &9.71		& 12.0 & --0.06	 $\pm$ 0.07	&Jul 26.62 &78.71 	& 2.8  & 0.11 $\pm$ 0.14  \\
May 20.82 &11.92	& 7.4  & --0.03 $\pm$ 0.11  	&Aug 4.23  &87.33 	& 4.1  & 0.43 $\pm$ 0.17 \\
May 21.61 &12.70  	& 12.0 & --0.08 $\pm$ 0.08  	&Aug 14.20 &97.29	& 3.6  & 0.22 $\pm$ 0.21 \\
May 23.43 &14.53	& 3.9  & --0.29 $\pm$ 0.13  	&Aug 16.25 &99.35	& 6.5  & 0.14 $\pm$ 0.10 \\
May 25.77 &16.87	& 3.3  & 0.11 $\pm$ 0.16  	&Aug 17.28 &100.38  	& 8.0  & 0.17 $\pm$ 0.08 \\
May 27.73 &18.83 	& 5.3  & 0.20 $\pm$ 0.11 	&Aug 18.23 &101.33	& 5.6  & 0.08 $\pm$ 0.10\\
May 30.39 &21.49	& 3.0  & 0.07 $\pm$ 0.13       	&Aug 20.30 &103.40 	& 5.6  & 0.25 $\pm$ 0.08\\
May 31.40 &22.50    	& 3.4  & 0.13 $\pm$ 0.12       	&Aug 22.23 &105.33	& 6.1  & 0.20 $\pm$ 0.09 \\   
Jun 1.39  &23.48	& 2.9  & 0.06 $\pm$ 0.14       	&Aug 29.17 &112.27	& 4.3  & 0.48 $\pm$ 0.12 \\ 
Jun 2.39  &24.49	& 3.4  & 0.10  $\pm$ 0.12       &Sep 1.17  &115.27	& 4.6  & 0.08$\pm$ 0.10 \\
Jun 6.42  &28.52	& 5.3  & 0.07 $\pm$ 0.15      	&Sep 9.22  &123.32	& 6.6  & 0.36  $\pm$ 0.09 \\
Jun 8.43  &30.52 	& 5.9  & 0.03 $\pm$ 0.37  	&Sep 23.22 &138.32 	& 10.2 & 0.26 $\pm$ 0.10 \\   
Jun 10.7  &32.81	& 4.1  & 0.05 $\pm$ 0.14	\\
\tableline
\end{tabular}
\end{flushleft}
\end{minipage}
\end{table}

\begin{table}
\caption[c]{1.4, 4.86 and 8.46 GHz long-term averages and 1.4 GHz 
spectral indices (with
respect to 8.46 and 4.86 GHz). The 8.46 and 4.86 GHz averages have been
determined from observations by Frail et al. (1997b). \label{tab:hist}}
\begin{minipage}{8.8cm}
\begin{flushleft}
\begin{tabular}{lrcccc}
\tableline
UT day (1997) & $F_{1.4\,
{\rm GHz}}$ & $F_{4.86\,{\rm GHz}}$ & $F_{8.46\,{\rm GHz}}$ &
$\alpha_{\rm 1.4-4.86 GHz}$ & 
$\alpha_{\rm 1.4-8.46 GHz}$ \\
\tableline
May  9.70 - May 16.63   	& 35 $\pm$ 40 & 330 $\pm$ 33 & 531
$\pm$ 12 & $>$ 1.05 & $>$ 1.04 \\
May 18.61- May 25.77    	& --70 $\pm$ 55 & 390 $\pm$ 25 & 669
$\pm$ 17 & $>$ 0.96 & $>$ 0.99\\
May 27.73-Jun 2.39		& 110 $\pm$ 70 & 728 $\pm$ 20 & 810
$\pm$ 14 & $>$ 1.30 & $>$ 0.97\\
Jun 6.42-Jun 16.44\footnote{Excluding June 8.43 UT}
& 55 $\pm$ 65 & 655 $\pm$ 21 & 612 $\pm$ 33 & $>$ 1.27 & $>$ 0.83\\
Jul 1.49 - Aug 4.23 		& 306 $\pm$ 65 	& 558 $\pm$ 13	& 594 
$\pm$ 14	& 0.48 $\pm$ 0.17 	& 0.37 $\pm$ 0.12 \\
\tableline
\end{tabular}
\end{flushleft}
\end{minipage}
\end{table}

\end{document}